\documentclass[onecolumn,usenatbib]{mn2e}
\usepackage{amssymb}
\usepackage{subfigure}
\usepackage{graphicx,natbib,rotating}

\title[The bulk kinetic power of radio jets in AGNs] {The bulk kinetic power of radio jets in active galactic
nuclei}
\author[M. F. Gu, X. Cao and D. R. Jiang]
{Minfeng Gu $^{1}$\thanks{E-mail: gumf@shao.ac.cn}, Xinwu Cao$^{1}$,
D. R. Jiang$^{1}$
\\
$^{1}$ Key Laboratory for Research in Galaxies and Cosmology,
Shanghai Astronomical Observatory, Chinese Academy of Sciences, 80\\
Nandan Road, Shanghai 200030, China}

\begin{document}
\pagerange{\pageref{firstpage}--\pageref{lastpage}} \pubyear{...}
\maketitle \label{firstpage}
\begin{abstract}

Based on the K\"onigl's inhomogeneous jet model, we estimate the jet
parameters, such as bulk Lorentz factor $\Gamma$, viewing angle
$\theta$ and electron number density $n_{\rm e}$ from radio VLBI and
X-ray data for a sample of active galactic nuclei (AGNs) assuming
that the X-rays are from the jet rather than the intracluster gas.
The bulk kinetic power of jets is then calculated using the derived
jet parameters. We find a strong correlation
between the total luminosity of broad emission lines and the bulk
kinetic power of the jets.
This result supports the scenario that the accretion process are
tightly linked with the radio jets, though how the disk and jet are
coupled is not revealed by present correlation analysis. Moreover,
we find a significant correlation between the bulk kinetic power and
radio extended luminosity. This implies that the emission from the
radio lobes are closely related with the energy flux transported
through jets from the central part of AGNs.

\end{abstract}


\section{Introduction}
The formation of highly relativistic jets in active galactic nuclei
(AGNs) is one of the unsolved fundamental problems in astrophysics
(e.g. Meier et al. 2001). It has been assumed that jets are produced
close to the central black hole, involving power extraction from the
black hole spin (Blandford \& Znajek 1977; Macdonald \& Thorne 1982;
Thorne \& Blandford 1982) and/or from the accretion disk (Blandford
\& Payne 1982). Although the jet formation remains unclear, the
estimate of the jet power is of fundamental physical interest, since
it can be used to quantify the power emerging from the central
engine of the radio source. Recently, the prescriptions for AGN
feedback have been introduced into semi-analytic models of galaxy
formation, and both Bower et al. (2006) and Croton et al. (2006)
show that this feedback is able to solve the issue of the bright end
of the luminosity function, whilst simultaneously solving other
problems of galaxy formation models such as why the most massive
galaxies are so red. Although the form of the AGN feedback adopted
is very different in the two prescriptions, the relativistic ejecta
from the AGN is a conceivably important ingredient of AGN feedback.
Indeed, in clusters of galaxies containing powerful radio sources,
X-ray observations have revealed bubbles and cavities in the hot
intracluster medium, evacuated by the expanding radio source (e.g.
McNamara et al. 2000; Fabian et al. 2003). Recent studies showed
that the mechanical luminosity of radio sources are sufficient to
suppress cluster cooling flows (Best et al. 2006; Nusser, Silk \&
Babul 2006). To understand the interaction between the radio sources
and the surrounding medium, it is clearly important to estimate the
bulk kinetic power of radio jets, since the expanding radio sources
provides a direct way for the AGN output to be coupled to its
environment.

The relation between the jets and the accretion processes in active
galactic nuclei has been extensively explored by many authors and in
different ways. The strong correlations have been found between the
low-frequency radio and narrow-line luminosities of 3C radio sources
(Baum \& Heckman 1989; Rawlings et al. 1989; Saunders et al. 1989),
and also between the broad line and extended radio luminosity for
radio-loud quasars (e.g. Cao \& Jiang 2001).
The link between the jets and the accretion processes can also been
studied through exploring the relationship between luminosity in
line emission and kinetic power of jets in different scales
(Rawlings \& Saunders 1991; Celotti \& Fabian 1993; Falcke, Malkan
\& Biermann 1995; Wang, Luo \& Ho 2004). Rawlings \& Saunders (1991)
used the narrow-line luminosity as indicative of the accretion power
and estimated the power transported by the jet from the energy
content and lifetime of the radio lobes, finding a good correlation
between the two. Using radio data on very long-baseline
interferometry (VLBI) scales and the standard synchrotron
self-Compton (SSC) theory, Celotti \& Fabian (1993) estimated the
jet kinetic power to put constraints on the matter content of jets.
It offers some clues to understand the fundamental questions of the
mechanisms, such as the collimation and acceleration of jets.
Celotti, Padovani \& Ghisellini (1997, hereafter C97) explored the
relation of luminosity in broad emission lines with the kinetic
power of the jets for a sample of radio-loud AGNs. Their estimate of
the bulk kinetic power is based on the adoption of the SSC model
applied to the radio VLBI data and X-ray (or optical) fluxes.
Lacking more accurate information, the minimum $\Gamma$ for any
given  $\delta$ [i.e. $\Gamma=0.5(\delta+1/\delta)$] is used in the
derivation of bulk kinetic power for objects with $\delta>1$,
otherwise the $\Gamma$ is derived from an average $\delta$.
They found a suggestive hint of correlation between these two
luminosities which is in favour of a link between the accretion
process and the
jets. However, by re-estimating the luminosity in broad emission
lines on the sample of C97, Wang et al. (2004) argued that the jet
bulk kinetic power is significantly correlated with the disk
luminosity. Maraschi \& Tavecchio (2003, hereafter MT03) found that
the jet power is linearly proportional to the disk power for a
sample of blazars, for which the jet powers were estimated using
physical parameters determined from uniformly modeling their
spectral energy distributions. However, by studying a sample of
quasars from Wang et al. (2004), Punsly \& Tingay (2005) argued that
the bulk kinetic power and the bolometric luminosity are very weakly
correlated in radio-loud quasars that possess blazar cores.

In the framework of the relativistic beaming and the SSC model, the
physical quantities in the jets can be estimated using the VLBI
observations and the X-ray flux density. Marscher (1987) derived the
beaming parameters on the assumption of homogeneous spherical
emission plasma. Ghisellini et al. (1993) adopted Marscher's
approach and obtained the Doppler boosting factor $\delta $ for 105
sources. Moreover, Readhead (1994)
estimated the equipartition Doppler boosting factor $%
\delta _{\rm eq}$, assuming that the sources are in equipartition
between the energy of radiating particles and the magnetic field.
G\"uijosa and Daly (1996) derived the $\delta _{\rm eq}$ for the
same sample in Ghisellini et al. (1993). The variability Doppler
factor $\delta_{\rm var}$ is derived on the assumption that the
associated variability brightness temperature of total radio flux
density flares are caused by the relativistic jets (L\"ahteenm\"aki
\& Valtaoja 1999). The advantage of homogeneous sphere model is that
the formalism is simple and the value of $\delta $ derived is
independent on the cosmology model. However, it is generally
difficult to know the component angular size and the flux at the
turnover frequency, so one has to assume that the VLBI observing
frequency is the synchrotron self-absorption frequency. In addition,
the dependence of core size on the observing frequency in some
sources is inconsistent with the homogeneous spherical assumption.
Blandford and K\"onigl (1979) and K\"onigl (1981) presented an
inhomogeneous relativistic jet model, in which both the flat
spectrum characteristics of some AGNs and the dependence of the core
size on the observing frequency could be well explained. Based on
their model, a new approach has been proposed to derive the jet
parameters including bulk Lorentz factor $\Gamma$, viewing angle
$\theta$ and electron number density $n_{\rm e}$ in the jets (Jiang,
Cao \& Hong 1998, hereafter J98). The proper motion measurements on
the jets' components were adopted in their calculations. The
correlation between the brightness temperature in the source rest
frame and the derived Doppler factor suggested that the derived
values of beaming parameters are quite reliable (J98). Moreover, the
derived beaming parameters from the homogeneous sphere model is in
general consistent with that from their inhomogeneous jet model.

In this work, we follow the method of J98 to derive the physical
quantities of jets for a large sample of AGNs, then re-analyze the
relation between the luminosity in broad line emissions and the bulk
kinetic power of the jets. In Section 2, we describe the sample of
sources. The method of jet parameters derivation and the estimate of
jet kinetic power are outlined in Section 3. Section 4 includes the
results and discussion. In the last section, we draw our
conclusions. The cosmology with $H_{0}=70 \rm {~km ~s^
{-1}~Mpc^{-1}}$, $\rm \Omega_{M}=0.3$, and $\rm \Omega_{\Lambda} =
0.7$ have been adopted throughout the paper.

\section{The sample}

In order to use inhomogeneous jet model to estimate the jet
parameters, all sources should have VLBI measurements of proper
motion of outflowing plasma. Combining with the relevant data of
sources, such as the radio flux density, the size of the core and
X-ray flux density, the jet parameters can be derived (J98). After
searching the literature, our sample is constructed, which consists
of 128 sources, including 94 quasars, 26 BL Lac objects and 8 radio
galaxies. The observational data for the sample are
presented in Table 1: 
(1) IAU name; (2) classification of the source (Q= quasars; Qc=
core-dominated quasars; Ql= lobe-dominated quasars; Qp= GHz peaked
quasars; BL= BL Lac objects; G= radio galaxies); (3) redshift z; (4)
observation frequency $\nu_{\rm s}$ in GHz; (5) core radio flux
density $f_{\rm c}$ at frequency $\nu_{\rm s}$; (6) VLBI core size
$\theta_{\rm d}$ in mas; (7) reference for the VLBI data; (8) the
proper motion $\mu_{\rm app}$; (9) reference for the proper motion;
(10) 1 keV X-ray flux density $f_{\rm 1keV}$ in $\mu$Jy; (11)
reference for the X-ray flux.

When there are more than one moving components, we adopted the
fastest one, which is regarded as a good approximation of jet bulk
motion. In addition, we use the core flux density measured at the
highest frequency, when VLBI core was measured at more than one
frequency. We assume that all the observed X-ray flux density is
attributable to the SSC emission in the derivation, which will
introduce some uncertainties. However, the derived jet parameters
are not sensitive to the adopted X-ray flux density (J98). The
redshift of 0716+714 is not available, and a value of 0.3 is assumed
in the calculation. To calculate the total luminosity of broad
emission lines, the available measurements of various broad
emission lines for each source are collected from literatures.
Moreover, we search the literatures and collect all available radio
extended emission data from VLA observations for each source. The
data are available for all 128 sources, of which the data from
Australia Telescope Compact Array (ATCA) are used for the southern
source 0208-512, and only upper limit is available for 8 sources due
to the faintness or non-detection of extended emission.
The extended flux density is k-corrected to 5 GHz in the rest frame
of the source assuming $\alpha_{\rm e}=-1$ ($f_{\rm ext}\propto
\nu^{\alpha_{\rm e}}$).

\section{Bulk kinetic power}

We estimate the bulk kinetic power based on the inhomogeneous jet
model (K\"onigl 1981). Using VLBI radio data, including proper
motion of plasma, and X-ray fluxes, we can calculate the comoving
electron number density $n_{\rm e}$, the magnetic field intensity,
Lorentz factor $\Gamma$ and the viewing angle $\theta$. A brief
description of the method is given below, and we refer to J98 and
references therein for a complete description. In brief, the jet
parameters were calculated by relating the model predicted size of
optically think region in the jet, radio emission from the optically
think region along the jet, SSC X-ray emission from the unresolved
jet, and the apparent transverse velocity to the observables of
radio core size, radio core flux density, X-ray flux density, and
the proper motion. In this paper, we assume that the X-rays are from
the jet rather than the intracluster gas and the radio blob speed is
the jet flow speed.

In K\"onigl's inhomogeneous jet model, the magnetic
field $B(r)$ and the number density of the relativistic electrons ${n_{\rm e}}(r,{%
\gamma _{\rm e}})$ in the jet are assumed to vary with the distance
from the apex
of the jet $r$ as $B(r)={B_1}(r/{r_1})^{-m}$ and ${n_{\rm e}}(r,{\gamma _{\rm e}})={n_1}%
(r/{r_1})^{-n}{\gamma _{\rm e}}^{-(2\alpha +1)}$, respectively,
where ${r_1}=1$ pc and $\gamma_{\rm e}$ is the Lorentz factor of the
electron in the jet. Given that the bulk motion velocity of the jet
is ${\beta }c$ (corresponding to a Lorentz factor $\Gamma$) with an
opening half-angle $\phi $, and the axis of the jet makes an angle
$\theta $ with the direction of the observer, the distance from the
origin of the jet, $r({\tau _{\nu _{\rm s}}}=1)$, at which the
optical depth to the synchrotron self-absorption at the observing
frequency $\nu _{\rm s}$ equals unity, is given as

\begin{equation}
{\frac{{r({\tau_{\nu_{\rm s}}}=
1)}}{{r_1}}}=(2c_{2}(\alpha)r_{1}n_{1}\phi\csc
\theta)^{2/(2\alpha+5)k_{\rm
m}}(B_{1}\delta)^{(2\alpha+3)/(2\alpha+5)k_{\rm m}}(\nu_{\rm s}
(1+z))^{-1/k_{\rm m}}
\end{equation}
where $c_{2}(\alpha)$ is the constant in the synchrotron absorption
coefficient, $\delta$ is the Doppler factor, and $k_{\rm
m}=[2n+m(2\alpha+3)-2] /(2\alpha+5)$.

The projection of the optically thick region in the jet is then used
as the observed VLBI core angular size $\theta_{\rm d}$,

\begin{equation}
\theta_{\rm d}= {\frac{{r({\tau_{\nu_{\rm s}}}= 1)\sin \theta}}{{\
D_{\rm a}}}}
\end{equation}
where $D_{\rm a}$ is the angular diameter distance of the source.

By integrating the emission from the optically thick region along
the jet, the radio flux of the core can be obtained

\begin{equation}
s(\nu_{\rm s})={\frac{{r_{1}^{2}\phi\sin\theta}}{{(4+m)\pi D_{\rm
a}^{2}}}} {\frac{{\ c_{1}(\alpha)}}{{c_{2}(\alpha)}}}
B_{1}^{-1/2}\nu_{\rm s}^{5/2} \left({\frac{
\delta}{{1+z}}}\right)^{1/2}\left({\frac{{r({\tau_{\nu_{\rm s}}}=
1)}}{{r_1}}} \right)^{(4+m)/2}
\end{equation}
where $\nu _{\rm s}$ is the VLBI observing frequency, and
${c_1}(\alpha )$ and ${c_2}(\alpha )$ are the constants in the
synchrotron emission and absorption coefficients, respectively.

Equation (13) in K\"onigl's work gives the X-ray flux density
estimation from an unresolved jet. As in J98, we adopt the
expression in the frequency region ${\nu _{\rm c}}>{\nu _{\rm
cb}}(r_{\rm M})$, where $r_{\rm M}$ is the smallest radius from
which optically thin synchrotron emission with spectral index
$\alpha $ is observed (K\"onigl 1981).

The proper motion observed with VLBI can be converted to the
apparent transverse velocity $\beta _{\rm app}$, which is related to
the bulk velocity of the jet ${\beta }c$ and viewing angle $\theta
$,

\begin{equation}
\beta_{\rm app}={\frac{{\beta\sin\theta}}{{1-\beta\cos\theta}}}
\end{equation}

Given the three parameters $\alpha $, $m$, $n$, and the relation
between the opening half angle $\phi $ and the Lorentz factor
$\Gamma$, the parameters of an inhomogeneous jet can be derived from
VLBI and X-ray observations, using the above equations and equation
(13) in K\"onigl (1981). In our calculation, we take $\alpha=0.75$,
the opening half-angle $\phi=1/\Gamma $, and assume $m=1$, $n=2$
corresponding to a free jet (Hutter \& Mufson 1986).

With the estimated comoving total electron number density $n_{\rm
t}$, Lorentz factor $\Gamma$ and the cross section of the jet $S$,
the bulk kinetic power is then derived as

\begin{equation}
L_{\rm kin}= Sn_{\rm t}(m_{\rm
e}\langle\gamma\rangle+m_{+}\langle\gamma_{+}\rangle)c^{2}\Gamma(\Gamma-1)\beta
c,
\end{equation}
where $m_{\rm e}$ is the electron rest mass, $m_{+}$ is the rest
mass of positive charge, $\langle\gamma\rangle$ is the average
Lorentz factor of electrons, and $\langle\gamma_{+}\rangle$ is the
average Lorentz factor of positive charges. For a conical jet with
an opening half-angle $\phi$, $S=2\pi r^{2}(1-\cos \phi)$. The total
electron number density $n_{\rm t}$ is given by

\begin{equation}
n_{\rm t}=\int\limits_{\gamma_{\rm min}}^{\gamma_{\rm max}}n_{\rm
e}(r, \gamma_{\rm e}) d\gamma_{\rm e},
\end{equation}

The bulk kinetic power of the inhomogeneous jet becomes

\begin{equation}
L_{\rm kin} = \alpha^{-1}\pi r^{2-n}r_{1}^{n}n_{1}\gamma_{\rm
min}^{-2\alpha} (1-\cos {\phi})(m_{\rm
e}\langle\gamma\rangle+m_{+}\langle\gamma_{+}\rangle)\Gamma(\Gamma-1)\beta
c^{3}
\end{equation}
With our adoption of $\alpha=0.75$, $\phi=1/\Gamma $, $m=1$, and
$n=2$, the bulk kinetic power of the jet is then given by


\begin{equation}
L_{\rm kin} = \frac{4}{3}\pi r_{1}^{2}n_{1}\gamma_{\rm
min}^{-\frac{3}{2}} (1-\cos {1/\Gamma})(m_{\rm
e}\langle\gamma\rangle+m_{+}\langle\gamma_{+}\rangle)\Gamma(\Gamma-1)\beta
c^{3}
\end{equation}
We note that the bulk kinetic power $L_{\rm kin}$ is independent of
$r$, since $n=2$ is adopted in the calculation and the particle
conservation is then satisfied along $r$. $L_{\rm kin}$ is largely
dependent of the matter content of jets and the low energy cut-off
$\gamma_{\rm min}$ of electrons. In present, the jet composition is
still unclear, i.e. whether electron-positron or electron-proton
(see Worrall \& Birkinshaw 2006, for a recent review and reference
therein). However, for an electron-proton plasma, $\gamma_{\rm
min}\sim 100$ has been suggested, while $\gamma_{\rm min}$ could be
as low as unity for an electron-positron jet (e.g. Celotti \& Fabian
1993). The detection of circular polarization strongly suggests that
the jets are electron-positron plasmas with $\gamma_{\rm min}
\lesssim 10$ at least in some sources (e.g. Wardle et al. 1998). The
similar conclusion is also arrived from powerful large scale X-ray
jets, if they are interpreted as inverse-Compton scattering of
cosmological microwave background photons in fast jets. From
equation (8), however, we find that the bulk kinetic power $L_{\rm
kin}$ for a electron-proton jet with $\gamma_{\rm min}\sim 100$
($m_{+}=1836~m_{\rm e}$ and $\langle\gamma_{+}\rangle=1$) is in
agreement with that of electron-positron one with $\gamma_{\rm min}
\lesssim 10$ ($m_{+}=m_{\rm e}$ and
$\langle\gamma_{+}\rangle=\langle\gamma\rangle$) within a factor
of three. 
In present work, we calculate the bulk kinetic power $L_{\rm kin}$
assuming electron-positron jets with $\gamma_{\rm min}=1$. A change
to $\gamma_{\rm min}=10$ will uniformly reduce $L_{\rm kin}$ by
about a factor of three. However, in this work, we mainly focus on
the correlation between the bulk kinetic power $L_{\rm kin}$ and the
luminosity in broad emission lines $L_{\rm BLR}$, therefore, the
value of $\gamma_{\rm min}$ will not affect the correlation
analysis, if the assumption that all sources have the same value
$\gamma_{\rm min}$ holds.

The observational data necessary for calculations are presented in
Table 1. 
Following C97, we use the line ratios reported by Francis et al.
(1991) and add the contribution from line $\rm H_{\alpha}$ to derive
the total broad line luminosity
$L_{\rm BLR}$. 
The derived values of jet parameters, $L_{\rm kin}$, $L_{\rm BLR}$
and 5 GHz radio extended luminosity for our sources are listed in
Table 2: (1) IAU name; (2) the viewing angle of jet $\theta$; (3)
the Lorentz factor $\Gamma$; (4) the Doppler factor $\delta$; (5)
the normalization factor of electron energy distribution $n_{1}$;
(6) the bulk kinetic power of jet $L_{\rm kin}$; (7) the total
luminosity in broad emission lines $L_{\rm BLR}$; (8) the references
for flux of broad emission lines used to estimate $L_{\rm BLR}$; (9)
the radio extended 5 GHz luminosity $L_{\rm ext, 5 GHz}$; (10) the
references for the radio extended flux.

\section{Results and discussion}

\subsection{Bulk kinetic power and BLR luminosity}

Out of 128 sources, the measurements of various broad emission lines
are only available for 98 sources from literature or Sloan Digital
Sky Survey (SDSS) spectra, including 81 quasars, 15 BL Lac objects,
and 2 radio galaxies. The reason of no measurements for remaining 30
sources could be either the non-detection of broad emission lines in
11 BL Lac objects and 6 radio galaxies, or no published line flux
measurements for 13 quasars. We have calculated the total luminosity
of broad emission lines for these 98 sources. The relationship
between $L_{\rm BLR}$ and $L_{\rm kin}$ is shown in Fig. 1. We find
a strong correlation between these two luminosities with a Spearman
correlation coefficient of $r=0.565$ at $\gg99.99\%$ confidence. It
should be noted that this correlation may be caused by the common
dependence on redshift. We present the bulk kinetic power and BLR
luminosity as functions of redshift $z$ for the sample in Fig. 2. We
therefore use the partial Spearman rank correlation method (Macklin
1982) to check this correlation. Still, a significant correlation
with a correlation coefficient of 0.323 is present at about $99.9\%$
significance level between $L_{\rm BLR}$ and $L_{\rm kin}$,
independent of the redshift.
We also perform a statistic analysis on the sources in the
restricted redshift range $0.5<z<1.0$. For this subsample of
sources, we check the correlation between luminosity and redshift,
and no correlation between either the bulk kinetic power or the BLR
luminosity and redshift, is found (see Fig. 2), while a significant
correlation is still present at 99.1 per cent confidence between the
bulk kinetic power and total broad-line luminosity (see Fig. 1).
Therefore, we conclude that this correlation might be intrinsic, at
least for our present sample. Assuming that the BLR luminosity is
due to the reprocessing of the ionizing radiation from the accretion
disk, it therefore strongly supports the scenario of a tight
connection between the relativistic jet and the accretion process.

For all 98 sources, the ordinary least-squares (OLS) bisector method
gives the following fit in Fig. 1:
\begin{equation}
\rm log~ \it L_{\rm kin}=\rm (0.86\pm0.07)~\rm log~ \it L_{\rm
BLR}+\rm (8.78\pm3.05)
\end{equation}
Apart from finding a significant correlation, Rawlings \& Saunders
(1991) found that the relationship between the bulk kinetic power
and narrow line luminosity is close to proportionality, $Q\propto
L_{\rm NLR}^{0.9\pm0.2}$, which extends over four orders of
magnitude. Our relation of $L_{\rm kin}\propto L_{\rm
BLR}^{0.86\pm0.07}$ is consistent with the relationship between the
jet power and narrow-line luminosity in Rawlings \& Saunders (1991)
and Celotti \& Fabian (1993), whereas it is somehow deviated, but
not much, from the linear relation between the jet power and disk
power found by MT03. However, it is much steeper than that of Wang
et al. (2004), $L_{\rm kin}\propto L_{\rm BLR}^{0.37}$. Although
what cause these differences is unclear, we note that the methods
used to estimate the jet power in MT03 and Wang et al. (2004) are
different from ours. The jet powers of MT03 were estimated using
physical parameters determined from uniformly modeling their
spectral energy distributions, while Wang et al. (2004) directly
used the jet bulk kinetic power from C97, which was estimated using
the homogeneous sphere SSC model. Moreover, MT03 obtained their disk
luminosities either directly from the optical-UV luminosity of the
big blue bump or from the original prescription of C97. The method
Wang et al. used to estimate $L_{\rm BLR}$ is basically same as
ours. In addition, Wang et al. sample consists of 35 blazars, and
only 16 sources (11 quasars and 5 BL Lac objects) were considered in
MT03. Their samples are much smaller than our sample. Whether these
factors influence the dependence of $L_{\rm kin}$ on $L_{\rm BLR}$
needs further investigations. Despite this, our results strongly
support the scenario that the accretion process are tightly linked
with the kinetic power in the jet, though how the disk and jet are
coupled is not revealed by present correlation analysis.



In general, BL Lac objects, which is thought to be FR I radio
galaxies pointing at us, are characterized by very weak or absent
emission lines, invisible blue bumps, and relatively powerful jets.
From Fig. 1, it is clear that BL Lac objects have fainter broad-line
luminosity compared to quasars, though only 15 BL Lac objects are in
our sample. We find that the $L_{\rm kin}-L_{\rm BLR}$ relation of
BL Lac objects deviate from that of quasars, although it generally
follows that of the whole sample. The linear fit using OLS bisector
method for BL Lac objects shows
\begin{equation}
\rm log~ \it L_{\rm kin}=\rm (0.71\pm0.12)~\rm log~ \it L_{\rm
BLR}+\rm (15.85\pm4.99)
\end{equation}
while for quasars, we have
\begin{equation} \rm log~ \it L_{\rm
kin}=\rm (1.12\pm0.08)~\rm log~ \it L_{\rm BLR}-\rm (2.94\pm3.39)
\end{equation}
Although the mechanism of jet formation is unclear, the different
dependence of $L_{\rm kin}$ on $L_{\rm BLR}$ in BL Lac objects and
quasars can be due to the difference of the accretion power as
measured in units of the Eddington one. Compared to quasars, BL Lac
objects are characterized by radiatively inefficient accretion disks
(Cao 2003), thus in these sources the jet power may be relatively
dominant. To further check the $L_{\rm kin}-L_{\rm BLR}$ correlation
in Fig. 1, we re-examine it for quasars only. When BL Lac objects
and radio galaxies are excluded, we still find a strong correlation
between $L_{\rm BLR}$ and $L_{\rm kin}$ with correlation coefficient
of $r=0.380$ at $99.95\%$ confidence. This further confirms the
tight link between the accretion process and the kinetic power in
the jet.

If the accretion process and the jet formation are indeed closely
related, then the tight relation between the mass channelled into
jets and that accreted by black hole would be expected. We then
investigate the relationship between the mass outflowing rate and
accretion one, on assumption of $L_{\rm bol}\approx10L_{\rm BLR}$
(Netzer 1990), and
with the expression of the kinetic and the accretion powers as
\begin{equation}
L_{\rm kin}=\Gamma \dot{M}_{\rm out}c^{2};~ L_{\rm
bol}=\eta\dot{M}_{\rm in}c^{2},
\end{equation}
where $\dot{M}_{\rm out}$ is the mass outflowing rate, $\Gamma$ is
jet Lorentz factor, $\dot{M}_{\rm in}$ is the mass accretion rate,
and $\eta$ is the efficiency of mass to energy conversion for
accretion. 
Adopting the typical value of $\eta\sim0.1$, we find a significant
correlation between $\dot{M}_{\rm out}$ and $\dot{M}_{\rm in}$ for
whole sample, with a Spearman correlation coefficient 0.514 at
$\gg99.99\%$ confidence. This implies that the mass outflowing rate
in jet is closely linked with the accretion one in accretion
disk. 

The present analysis is based on the derivation of jet parameters
using inhomogeneous jet model.
Some parameters and assumptions are adopted in the inhomogeneous jet
model to derive the physical quantities of the jet (J98), which may
induce some uncertainties in the estimation of $L_{\rm kin}$. The
most important is probably the intrinsic differences of the low
energy cut-off $\gamma_{\rm min}$ of electrons between the radio
sources themselves. In this work, we adopt the same value of
$\gamma_{\rm min}$ in deriving the kinetic power of the jet, which
may not be true. 
However, we are not able to estimate $\gamma_{\rm min}$ for each
source at present stage. Nevertheless, we believe that the adoption
of an uniform $\gamma_{\rm min}$ for all sources in the correlation
analysis would not affect the
main conclusion drawn here. 
Moreover, our results are based on the assumption of $\alpha=0.75$,
$m=1$, and $n=2$, and we adopted these same values for all sources
in our model calculation. However, we find that the alternative
adoption of $\alpha$, $m$, and $n$ do not change our main
conclusion, e.g. the strong correlation between $L_{\rm kin}$ and
$L_{\rm BLR}$. In practice, the sources may have different values of
parameters $\alpha $, $m$, and $n$, and, in principle, these
parameters could be constrained by the observable quantities (J98).
Unfortunately, the information is only found for a few cases through
multi-frequencies VLBI observations. Further high resolution
multi-frequencies VLBI observations would be helpful to improve our
model calculations.

It should be noted that not all sources in our sample have available
BLR luminosity. Therefore, the selection effects may be introduced
in our correlation analysis, i.e. those sources without published
broad line flux measurements may likely be biased towards those with
weak lines, especially the 11 BL Lac objects without broad line flux
measurements. To evaluate the selection effects, we tentatively
calculate the upper limit of BLR luminosity assuming equivalent
width of broad $\rm H\beta$ line $\rm EW<5\AA$ for these 11 BL Lacs.
Moreover, we calculate the BLR luminosity for 13 quasar without BLR
luminosity by randomly assigning the BLR flux in the BLR flux range
of 81 quasars with BLR luminosity. Combining these 24 sources with
those having BLR luminosity, we use the Astronomy Survival Analysis
(ASURV) package (Isobe, Feigelson \& Nelson 1986) to investigate the
correlation by taking the upper limit into account. A significant
correlation is still found with Spearman's rho correlation method.
This correlation is confirmed by using the partial correlation
method for censored data of Akritas \& Siebert (1996) to exclude the
common dependence of redshift. Furthermore, the correlation remains
significant even we conservatively adopt broad $\rm H\beta$ line
$\rm EW<1\AA$ for 11 BL Lacs. It thus seems that the non-BLR
luminosity sources do not affect our correlation results.

\subsection{Bulk kinetic power and radio extended luminosity}

In Fig. 3, the relation between bulk kinetic power and 5 GHz radio
extended luminosity is shown for all 128 sources. Since only the
upper limit of 5 GHz extended luminosity is given for 8 sources,
we use ASURV package (Isobe et al. 1986) to investigate the
correlation and perform the linear regression analysis for our
censored data. We find the significant correlation with a Spearman's
rho correlation coefficient of $r=0.493$ at $\gg99.99\%$ confidence.
Using the partial correlation method for censored data of Akritas \&
Siebert (1996) to exclude the common dependence of redshift, a
significant correlation is still present between $L_{\rm ext,5GHz}$
and $L_{\rm kin}$. We use the Schmitt-binning method (Schmitt 1985)
to perform $y/x$ and $x/y$ fits and then calculate a bisector of
these two fits, as described in Shapley, Fabbiano \& Eskridge (2001)
(see also Isobe et al. 1990). We obtain:
\begin{equation}
\rm log~ \it L_{\rm kin}=\rm (0.82\pm0.09)~\rm log~ \it L_{\rm
ext,5GHz}+\rm (12.38\pm4.00)
\end{equation}
which is shown as the solid line in Fig. 3.

The extended radio flux is usually emerged from the optically thin
radio lobes, and thus is free from the Doppler boosting effects,
since the lobe material is generally thought to be of low enough
bulk velocity. Therefore, the extended radio luminosity can be a
good tracer of jet power (e.g. Cao \& Jiang 2001). The significant
correlation between $L_{\rm kin}$ and $L_{\rm ext,5GHz}$ implies
that the emission from radio lobes are tightly related with the
energy ejected into the jet from the central parts of AGNs. This
result is not surprising, as it could be naturally expected.
Although the detailed mechanism of jet formation is still unclear,
the energy can be transported through the jets to the radio lobes
once the jets are generated. Most of the energy flux of jets is not
radiated away, instead are in mechanical form (i.e. bulk kinetic
power), of which the particles and fields are necessary to produce
the synchrotron luminosity that is detected in the radio lobes. If
the radiative efficiency of radio lobes are similar between our
radio sources, then the tight link between the bulk kinetic power of
jets and the radio extended emission is expected, since the latter
is optically thin and not effected from the Doppler enhancement.

Motivated largely by the observed effects of radio-loud AGN on their
environments at galaxy cluster scales (e.g. Fabian et al. 2003),
whether the heating effect of AGN activity, particularly radio-loud
AGN activity, can balance the cooling of the gas has recently arose
much interest (e.g. Best et al. 2005, 2006; Croton et al. 2006;
Nusser, Silk \& Babul 2006). Independently of the radio properties,
B$\hat{\rm i}$rzan et al. (2004) estimated the mechanical luminosity
associated with the radio source, by studying the cavities and
bubbles that are produced in clusters and groups of galaxies due to
the interactions between the radio sources and the surrounding hot
gas. The dependence of the mechanical luminosity on the 1.4 GHz
radio luminosity of the associated radio sources, $L_{\rm
mech}\propto L_{\rm radio}^{0.44\pm0.06}$ fitted for their entire
sample, is somewhat deviated from ours (equation 13). However, the
dependence for the radio-filled cavities only (see B$\hat{\rm
i}$rzan et al. 2004 for details), $L_{\rm mech}\propto L_{\rm
radio}^{0.6\pm0.1}$ is marginally consistent with ours within the
errors.
Despite this, we note that their work is mainly based on the galaxy
clusters and the radio sources in these clusters, however, our
present study focused on the powerful radio sources. Moreover, the
jet power is estimated using different methods, and the different
radio luminosity is used. In present, we are not able to draw a
solid calibration between the radio emission and the kinetic power
of jets, and it needs further investigations.


Despite the strong correlation presented in Fig. 3, the significant
scatter is clearly seen. This is not surprising since even for a
source of fixed jet kinetic power the radio luminosity changes as
the source ages (e.g. Kaiser et al. 1997). However, there are
several factors that can introduce the scatter into the correlation.
We note that the observed radio extended emission has been
dissipated over a long period, which is not contemporaneous with the
estimated bulk kinetic power. Moreover, when jets transported the
energy flux from the central parts of AGNs to outer radio lobes, the
jets can be decelerated by the interaction with the nuclear ISM
and/or the entrainment of external gas (e.g. Tavecchio et al. 2006).
As a result, the part of jet power will be lost to the ISM. In some
extreme cases, the kinetic power of the jet on kiloparsec scales
could be about three orders of magnitude weaker than the power of
the jet on 10 - 100 pc scales due to the jet-ISM interaction, i.e.
virtually all of the jet power can be lost to the ISM within the
inner kiloparsec (Gallimore et al. 2006). Consequently, the
difference of jet-ISM interactions between the radio themselves may
bring scatter. Furthermore, part of the scatter may be due to the
different radiative efficiency in individual source.

\subsection{Bulk kinetic power versus radiative luminosity}

It is well known that the monochromatic radio luminosity does not
provide a good indicator of the mechanical energy output of a radio
source. Radio sources are inefficient radiators. Bicknell (1995)
estimated that the kinetic energy output of a radio jet is typically
a factor of 100-1000 higher than the total radio luminosity of a
radio source, which is recently confirmed by the observations of
B$\hat{\rm i}$rzan et al. (2004).

We estimate the amount of radiative dissipation on parsec scales,
i.e. the ratio of $\it L_{\rm kin}/\it L_{\rm rad,in}$ of the bulk
kinetic power to the intrinsic radiative luminosity. The latter has
been computed from the observed VLBI radio core fluxes. The Doppler
correction on the monochromatic luminosity is assumed to be $L_{\rm
obs}=\delta ^{p}~L_{\rm int}$, where $L_{\rm obs}$ and $L_{\rm int}$
are the observed and intrinsic (comoving)
luminosities. 
We can calculate $p$ from the dependence of the core radio flux on
the Doppler factor from equation (3), in which $p$ is dependent of
the value of $\alpha$, $m$ and $n$. In Fig. 4, we show the histogram
of the derived ratios $\it L_{\rm kin}/\it L_{\rm rad,in}$. We find
that the ratio covers about three orders of magnitude with the
average value $<\rm log~(\it L_{\rm kin}/\it L_{\rm rad,in})>=\rm
4.98\pm0.79$. This result is consistent with that of Celotti \&
Fabian (1993), although a wider spread of ratio distribution in
their sources. The results indicate that for all sources the kinetic
power is dominant with respect to the radiative output, and
consequently that the radiative dissipation is not an efficient
process. Moreover, the large variation in this ratio indicates that
the radio luminosity is not necessarily a reliable probe of the
available bulk kinetic power.

It is commonly accepted that the synchrotron emission of blazars can
extend to optical and even X-ray region, which can dominate over the
thermal emission from accretion disk, and the radio emission as
well. The spectral energy distribution (SED) of blazars is usually
composed of two peaks, of which the first one is dedicated to the
synchrotron emission for jets, and the second is due to the inverse
Compton process (Fossati et al. 1998; Ghisellini et al. 1998). Thus
the total radiative luminosity of the jets, if we can estimate from
integrating over the synchrotron domain of SED, can represent the
minimum power that must be associated with the jet in order to
produce the observed luminosity. In this sense, the radio emission
solely might not be a good indicator of the radiative output from
the radio jets, i.e. the radiation losses of the kinematic jet flow.
MT03 found that the radiative efficiency, i.e. the ratio of the
total radiative luminosity of the jet to the jet power, can be in
the range 1\%-10\%. Even so, the radiative dissipation is still not
an efficient process, and the most of the energy flux is in the
kinetic form. In present, it is not readily to estimate the total
radiative luminosity of jets for our sample. Nevertheless, we
believe that the inefficient radiators of radio sources would be
still retained.


\section{Conclusions}

Based on the inhomogeneous jet model, we have calculated the jet
parameters for a sample of AGNs. The bulk kinetic power of radio
jets are then estimated using the derived jet parameters. We found a
significant correlation between the bulk kinetic power of the
relativistic jet and the total luminosity in broad emission lines,
implying a tight link between the jet and accretion process.
Moreover, the bulk kinetic power of jets are strongly correlated
with the radio extended luminosity. This indicates a closely
connection between the emission from radio lobes and the energy flux
transported through jets from the central parts of AGNs. In
addition, we found that the bulk kinetic power is dominant with
respect to the radiative output, which means the radiative
dissipation is not an efficient process.

\section*{Acknowledgments}

We thank the anonymous referee for insightful comments and
constructive suggestions. This work is supported by National Science
Foundation of China (grants 10633010, 10703009, 10833002, 10773020
and 10821302), 973 Program (No. 2009CB824800), and the CAS
(KJCX2-YW-T03). This research has made use of the NASA/ IPAC
Extragalactic Database (NED), which is operated by the Jet
Propulsion Laboratory, California Institute of Technology, under
contract with the National Aeronautics and Space Administration.

Funding for the SDSS and SDSS-II has been provided by the Alfred P.
Sloan Foundation, the Participating Institutions, the National
Science Foundation, the U.S. Department of Energy, the National
Aeronautics and Space Administration, the Japanese Monbukagakusho,
the Max Planck Society, and the Higher Education Funding Council for
England. The SDSS Web Site is http://www.sdss.org/.

The SDSS is managed by the Astrophysical Research Consortium for the
Participating Institutions. The Participating Institutions are the
American Museum of Natural History, Astrophysical Institute Potsdam,
University of Basel, University of Cambridge, Case Western Reserve
University, University of Chicago, Drexel University, Fermilab, the
Institute for Advanced Study, the Japan Participation Group, Johns
Hopkins University, the Joint Institute for Nuclear Astrophysics,
the Kavli Institute for Particle Astrophysics and Cosmology, the
Korean Scientist Group, the Chinese Academy of Sciences (LAMOST),
Los Alamos National Laboratory, the Max-Planck-Institute for
Astronomy (MPIA), the Max-Planck-Institute for Astrophysics (MPA),
New Mexico State University, Ohio State University, University of
Pittsburgh, University of Portsmouth, Princeton University, the
United States Naval Observatory, and the University of Washington.


{}

\clearpage

\begin{figure}
  \begin{center}
    \includegraphics[height=.35\textheight,width=.5\textwidth]{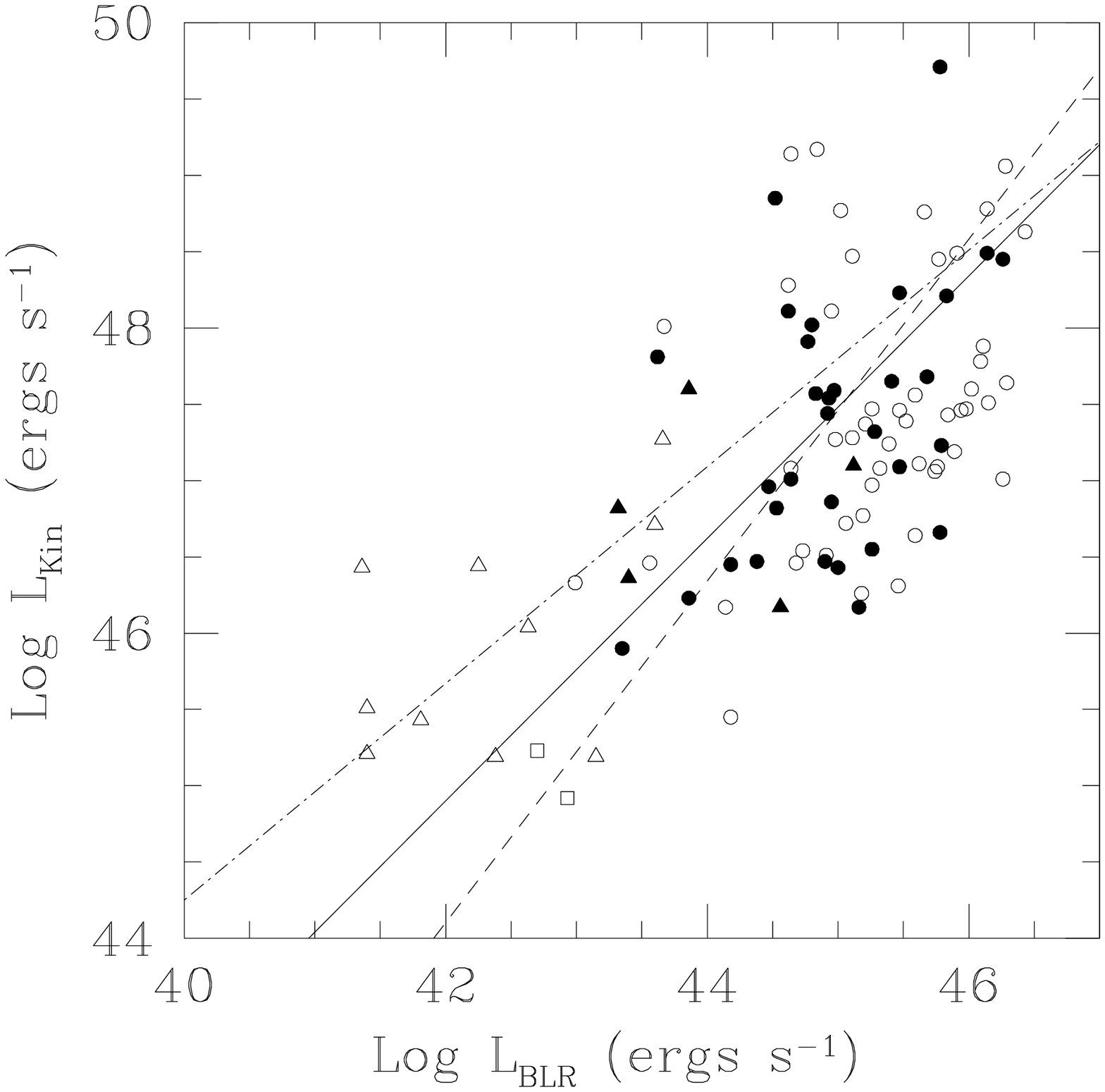}
  \end{center}
\caption{The bulk kinetic power versus BLR luminosity. The circles
represent quasars, and the triangles are BL Lac objects, while the
rectangles show radio galaxies. The filled symbols are the sources
in the redshift range $0.5<z<1.0$. The solid line is the fitted line
for the whole sample using the OLS bisector method, and the dashed
line is fitted for quasars only, while the dot-dashed line is fitted
for BL Lac objects only.}
\end{figure}

\begin{figure}
  \begin{center}
    \includegraphics[height=.35\textheight,width=.5\textwidth]{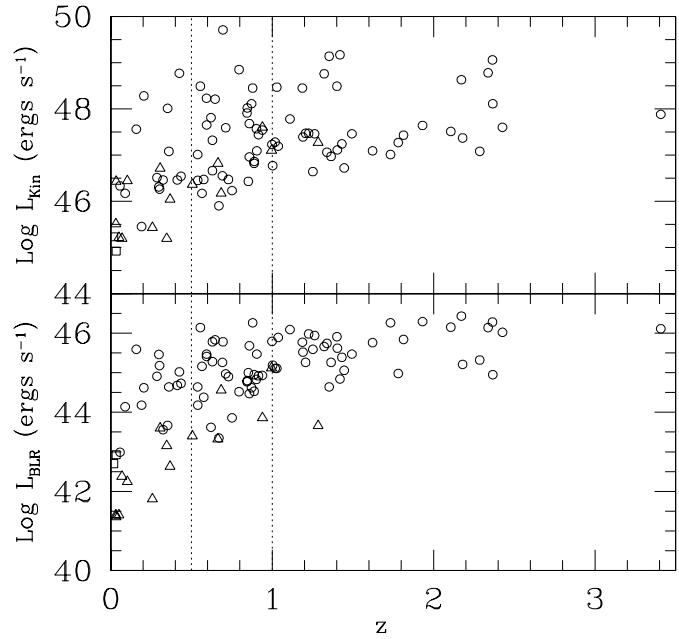}
  \end{center}
\caption{The bulk kinetic power versus redshift (upper panel) and
the BLR luminosity versus redshift planes (lower panel) for the
sample. The circles represent quasars, and the triangles are BL Lac
objects, while the rectangles show radio galaxies. The restricted
redshift range, $0.5<z<1.0$, is indicated with the dotted lines.}
\end{figure}

\begin{figure}
  \begin{center}
    \includegraphics[height=.35\textheight,width=.5\textwidth]{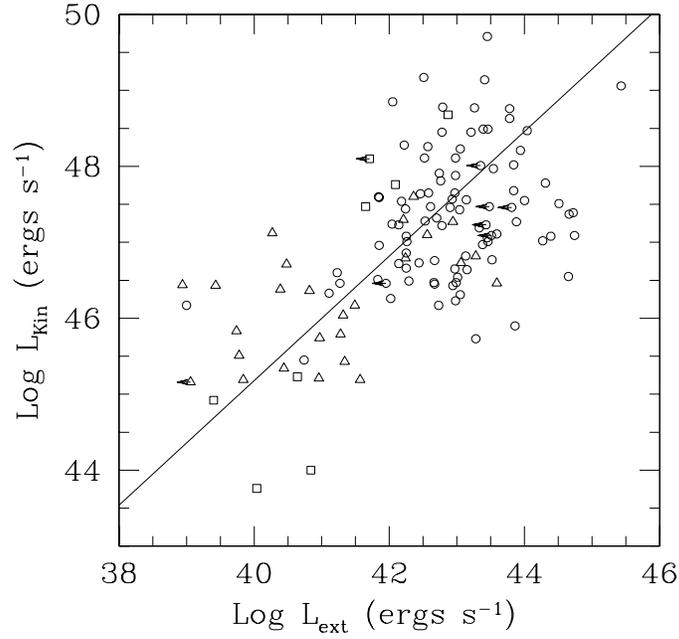}
  \end{center}
\caption{The bulk kinetic power versus radio
 5 GHz extended luminosity. The symbols are the same as in Fig. 2. The arrows indicate
 the upper limit of extended luminosity. The solid line is
 the bisector linear fit using Schmitt-binning method for censored data (see text for details).}
\end{figure}

\begin{figure}
  \begin{center}
    \includegraphics[height=.35\textheight,width=.5\textwidth]{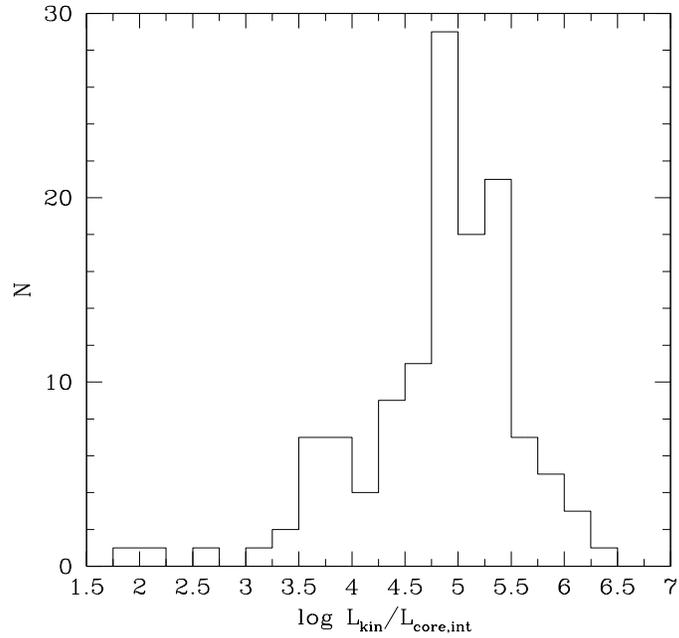}
  \end{center}
\caption{The histogram shows the distribution of the ratio of the
bulk kinetic power to intrinsic core radio radiative luminosities.}
\end{figure}

\clearpage

\begin{table*}\centering\begin{minipage}{140mm}
\caption{VLBI and X-ray Data of the Sample.}
\begin{tabular}{cccrccccrrr}
\hline\hline
 Source &
 Type & z & $\nu_{\rm s}$ & $f_{\rm c}$ & $\theta_{\rm
d}$ & Refs.
& $\mu_{\rm app}$ & Refs. & $f_{\rm 1keV}$ & Refs. \\
 &  &  & (GHz) &
(Jy) & (mas) &  & ($\rm mas~yr^{-1}$) &  & ($\rm \mu Jy$) &  \\
\hline
0003$-$066  &  BL  &  0.347  &  15.0  &  1.850  &   0.17  & 1   & 0.010  & 11  &  0.169 & 13  \\
0007$+$106  &  Qc  &  0.089  &  43.0  &  1.540  &   0.07  & 2   & 0.237  & 2   &  2.740 & 14  \\
0016$+$731  &  Qc  &  1.781  &  15.0  &  1.020  &   0.10  & 1   & 0.220  & 4   &  0.050 & 15  \\
0035$+$413  &  Qc  &  1.353  &  15.0  &  0.380  &   0.27  & 1   & 0.100  & 11  &  0.099 & 14  \\
0106$+$013  &  Qc  &  2.107  &  15.0  &  2.320  &   0.12  & 1   & 0.280  & 11  &  0.220 & 4   \\
0108$+$388  &  G   &  0.670  &   8.5  &  0.240  &   0.21  & 3   & 0.100  & 12  &  0.060 & 16  \\
0112$-$017  &  Qc  &  1.365  &  15.0  &  0.480  &   0.08  & 1   & 0.020  & 11  &  0.150 & 15  \\
0133$+$207  &  Ql  &  0.425  &  10.7  &  0.082  &   0.22  & 4   & 0.240  & 4   &  0.753 & 4   \\
0133$+$476  &  Qc  &  0.859  &  15.0  &  4.710  &   0.09  & 1   & 0.040  & 11  &  0.300 & 15  \\
0153$+$744  &  Qc  &  2.338  &  15.0  &  0.190  &   0.11  & 1   & 0.080  & 4   &  1.000 & 4   \\
0202$+$149  &  Qc  &  0.405  &  15.0  &  1.760  &   0.16  & 1   & 0.250  & 11  &  0.060 & 15  \\
0208$-$512  &  Qc  &  1.003  &   5.0  &  2.770  &   0.35  & 4   & 0.600  & 4   &  0.080 & 4   \\
0212$+$735  &  Qc  &  2.367  &  15.0  &  2.400  &   0.20  & 1   & 0.090  & 4   &  0.260 & 15  \\
0219$+$428  &  BL  &  0.444  &  43.2  &  0.593  &   0.02  & 5   & 1.110  & 5   &  1.560 & 15  \\
0234$+$285  &  Qc  &  1.207  &  22.3  &  1.700  &   0.10  & 4   & 0.300  & 4   &  0.150 & 4   \\
0235$+$164  &  BL  &  0.940  &   5.0  &  1.750  &   0.50  & 4   & 0.840  & 4   &  0.170 & 4   \\
0316$+$413  &  G   &  0.017  &  22.2  &  6.000  &   0.30  & 4   & 0.540  & 4   & 18.000 & 4   \\
0333$+$321  &  Qc  &  1.263  &  15.0  &  1.840  &   0.12  & 1   & 0.400  & 11  &  0.440 & 15  \\
0336$-$019  &  Qc  &  0.852  &  15.0  &  1.780  &   0.07  & 1   & 0.420  & 5   &  0.100 & 17  \\
0415$+$379  &  G   &  0.049  &  86.2  &  2.900  &   0.13  & 4   & 1.540  & 4   &  3.283 & 4   \\
0420$-$014  &  Qc  &  0.915  &  43.2  &  2.724  &   0.06  & 5   & 0.290  & 11  &  0.370 & 15  \\
0430$+$052  &  G   &  0.033  &  15.0  &  1.710  &   0.10  & 1   & 2.660  & 4   & 10.000 & 4   \\
0440$-$003  &  Qc  &  0.844  &  15.0  &  0.620  &   0.15  & 1   & 0.340  & 5   &  0.189 & 13  \\
0454$+$844  &  BL  &  0.112  &   5.0  &  1.300  &   0.55  & 4   & 0.140  & 4   &  0.050 & 4   \\
0458$-$020  &  Qc  &  2.286  &  43.2  &  0.934  &   0.02  & 5   & 0.150  & 4   &  0.100 & 17  \\
0528$+$134  &  Qc  &  2.060  &  43.2  &  3.875  &   0.07  & 5   & 0.400  & 4   &  0.310 & 15  \\
0552$+$398  &  Qp  &  2.365  &   8.4  &  2.620  &   0.73  & 6   & 0.040  & 12  &  0.490 & 6   \\
0605$-$085  &  Qc  &  0.872  &  15.0  &  1.790  &   0.27  & 1   & 0.180  & 11  &  0.168 & 18  \\
0607$-$157  &  Qc  &  0.324  &  15.0  &  6.920  &   0.19  & 1   & 0.170  & 11  &  0.290 & 14  \\
0615$+$820  &  Qc  &  0.710  &   5.0  &  0.610  &   0.50  & 4   & 0.050  & 4   &  0.040 & 15  \\
0642$+$449  &  Qp  &  3.408  &  15.0  &  2.920  &   0.12  & 1   & 0.010  & 11  &  0.120 & 15  \\
0710$+$439  &  G   &  0.518  &   5.0  &  0.630  &   0.96  & 4   & 0.040  & 4   &  0.550 & 4   \\
0716$+$714  &  BL  &  0.300  &  43.2  &  0.390  &   0.04  & 5   & 1.200  & 5   &  0.990 & 15  \\
0723$+$679  &  Ql  &  0.846  &  43.0  &  0.677  &   0.06  & 7   & 0.190  & 12  &  0.162 & 18  \\
0735$+$178  &  BL  &  0.424  &  15.0  &  0.950  &   0.14  & 1   & 0.640  & 11  &  0.220 & 15  \\
0736$+$017  &  Qc  &  0.191  &  15.0  &  1.450  &   0.06  & 1   & 0.930  & 11  &  0.640 & 15  \\
0738$+$313  &  Qc  &  0.630  &  15.0  &  0.870  &   0.11  & 1   & 0.070  & 11  &  0.075 & 14  \\
0745$+$241  &  Qc  &  0.409  &  15.0  &  0.830  &   0.10  & 1   & 0.320  & 11  &  0.131 & 19  \\
0748$+$126  &  Qc  &  0.889  &  15.0  &  2.860  &   0.11  & 1   & 0.274  & 11  &  0.209 & 20  \\
0754$+$100  &  BL  &  0.266  &  15.0  &  1.420  &   0.11  & 1   & 0.700  & 11  &  0.720 & 15  \\
0804$+$499  &  Qc  &  1.432  &  15.0  &  1.020  &   0.09  & 1   & 0.130  & 11  &  0.170 & 15  \\
0808$+$019  &  BL  &  0.930  &  15.0  &  1.270  &   0.04  & 1   & 0.110  & 11  &  0.380 & 15  \\
0814$+$425  &  BL  &  0.258  &  15.0  &  1.080  &   0.06  & 1   & 0.320  & 11  &  0.050 & 15  \\
0823$+$033  &  BL  &  0.506  &  15.0  &  1.100  &   0.07  & 1   & 0.480  & 11  &  0.415 & 13  \\
0827$+$243  &  Qc  &  0.939  &  43.2  &  1.406  &   0.05  & 5   & 0.480  & 5   &  0.340 & 17  \\
0829$+$046  &  BL  &  0.180  &  22.2  &  0.796  &   0.05  & 5   & 1.400  & 5   &  0.400 & 15  \\
0836$+$710  &  Qc  &  2.172  &  43.2  &  1.570  &   0.06  & 5   & 0.240  & 5   &  2.260 & 15  \\
0850$+$581  &  Qc  &  1.322  &  15.0  &  0.070  &   0.08  & 1   & 0.200  & 11  &  0.970 & 4   \\
0851$+$202  &  BL  &  0.306  &  43.2  &  1.640  &   0.04  & 5   & 0.670  & 5   &  2.240 & 15  \\
0859$-$140  &  Ql  &  1.339  &  15.0  &  1.170  &   0.09  & 1   & 0.260  & 11  &  0.171 & 14  \\
0906$+$015  &  Qc  &  1.018  &  15.0  &  2.360  &   0.15  & 1   & 0.220  & 11  &  0.141 & 14  \\
0906$+$430  &  Qc  &  0.670  &   5.0  &  0.875  &   0.10  & 4   & 0.180  & 4   &  0.090 & 4   \\
0917$+$449  &  Qc  &  2.180  &  22.2  &  1.042  &   0.05  & 5   & 0.150  & 5   &  0.470 & 15  \\
0917$+$624  &  Q   &  1.446  &   8.4  &  1.220  &   0.11  & 4   & 0.230  & 4   &  0.120 & 4   \\
0923$+$392  &  Qc  &  0.695  &  15.0  &  0.230  &   0.34  & 1   & 0.180  & 4   &  0.370 & 4   \\
0945$+$408  &  Qc  &  1.252  &  15.0  &  0.990  &   0.06  & 1   & 0.370  & 11  &  0.110 & 15  \\
0953$+$254  &  Qc  &  0.712  &  15.0  &  0.360  &   0.12  & 1   & 0.310  & 11  &  0.097 & 14  \\
0954$+$658  &  BL  &  0.368  &   5.0  &  0.477  &   0.19  & 4   & 0.440  & 4   &  0.160 & 15  \\
1012$+$232  &  Qc  &  0.565  &  15.0  &  1.080  &   0.07  & 1   & 0.270  & 11  &  0.088 & 14  \\
1015$+$359  &  Qc  &  1.226  &  15.0  &  0.710  &   0.13  & 1   & 0.200  & 11  &  0.051 & 14  \\
1039$+$811  &  Qc  &  1.260  &   8.5  &  0.450  &   0.09  & 8   & 0.070  & 12  &  0.180 & 15  \\
\hline
\end{tabular}
\end{minipage}
\end{table*}

\newpage
\addtocounter{table}{-1}
\begin{table*}\centering\begin{minipage}{140mm}
\caption{$Continued.$}
\begin{tabular}{cccrccccrrr}
\hline\hline
 Source &
 Type & z & $\nu_{\rm s}$ & $f_{\rm c}$ & $\theta_{\rm d}$ & Refs.
& $\mu_{\rm app}$ & Refs. & $f_{\rm 1keV}$ & Refs. \\
 &  &  & (GHz) &
(Jy) & (mas) &  & ($\rm mas~yr^{-1}$) &  & ($\rm \mu Jy$) &  \\
\hline
1040$+$123  &  Qc  &  1.029  &  10.7  &  0.590  &   0.33  & 4   & 0.110  & 4   &  0.121 & 4   \\
1049$+$215  &  Qc  &  1.300  &  15.0  &  1.220  &   0.16  & 1   & 0.140  & 11  &  0.064 & 14  \\
1055$+$018  &  Qc  &  0.888  &  15.0  &  4.930  &   0.06  & 1   & 0.040  & 11  &  0.210 & 15  \\
1055$+$201  &  Qc  &  1.110  &  15.0  &  0.260  &   0.10  & 1   & 0.180  & 11  &  0.164 & 20  \\
1101$+$384  &  BL  &  0.031  &  15.0  &  0.450  &   0.10  & 1   & 1.330  & 4   & 36.100 & 15  \\
1127$-$145  &  Qc  &  1.187  &  22.2  &  1.060  &   0.14  & 5   & 0.520  & 5   &  0.340 & 15  \\
1128$+$385  &  Qc  &  1.733  &  15.0  &  0.940  &   0.05  & 1   & 0.010  & 11  &  0.061 & 14  \\
1137$+$660  &  Ql  &  0.646  &   8.4  &  0.119  &   0.23  & 9   & 0.060  & 12  &  0.343 & 14  \\
1150$+$812  &  Qc  &  1.250  &   5.0  &  0.460  &   0.50  & 4   & 0.110  & 4   &  0.200 & 4   \\
1156$+$295  &  Qc  &  0.729  &  22.2  &  1.372  &   0.05  & 5   & 0.540  & 5   &  0.440 & 15  \\
1219$+$285  &  BL  &  0.102  &  22.2  &  0.263  &   0.09  & 5   & 0.600  & 5   &  0.400 & 15  \\
1222$+$216  &  Ql  &  0.435  &  22.2  &  0.960  &   0.06  & 5   & 0.900  & 5   &  0.410 & 17  \\
1226$+$023  &  Qc  &  0.158  &  43.2  &  8.040  &   0.13  & 5   & 1.600  & 5   & 20.420 & 15  \\
1228$+$127  &  G   &  0.004  &  15.0  &  1.390  &   0.33  & 1   & 3.070  & 4   &  0.680 & 4   \\
1253$-$055  &  Qc  &  0.538  &  43.2  & 13.773  &   0.07  & 5   & 0.310  & 5   &  1.500 & 15  \\
1302$-$102  &  Qc  &  0.286  &  15.0  &  0.530  &   0.09  & 1   & 0.310  & 11  &  0.723 & 14  \\
1308$+$326  &  BL  &  0.996  &  15.0  &  2.590  &   0.14  & 1   & 0.750  & 4   &  0.110 & 15  \\
1334$-$127  &  Qc  &  0.539  &  15.0  &  7.400  &   0.04  & 1   & 0.050  & 11  &  0.450 & 15  \\
1345$+$125  &  G   &  0.121  &   8.5  &  0.480  &   0.89  & 10  & 0.160  & 11  &  0.038 & 21  \\
1406$-$076  &  Q   &  1.494  &  22.2  &  0.833  &   0.07  & 5   & 0.630  & 5   &  0.075 & 14  \\
1413$+$135  &  BL  &  0.247  &  15.0  &  1.420  &   0.04  & 1   & 0.450  & 11  &  0.050 & 15  \\
1458$+$718  &  Qc  &  0.905  &   8.5  &  0.880  &   0.17  & 3   & 0.250  & 12  &  0.226 & 14  \\
1508$-$055  &  Ql  &  1.191  &  15.0  &  0.590  &   0.09  & 1   & 0.530  & 11  &  0.147 & 13  \\
1510$-$089  &  Qc  &  0.360  &  43.2  &  1.458  &   0.05  & 5   & 0.850  & 11  &  0.490 & 15  \\
1532$+$016  &  Qc  &  1.420  &  15.0  &  0.320  &   0.23  & 1   & 0.210  & 11  &  0.130 & 22  \\
1546$+$027  &  Qc  &  0.412  &  15.0  &  2.760  &   0.08  & 1   & 0.050  & 11  &  0.840 & 15  \\
1548$+$056  &  Qc  &  1.422  &  15.0  &  0.880  &   0.11  & 1   & 0.052  & 11  &  0.018 & 13  \\
1606$+$106  &  Qc  &  1.226  &  15.0  &  1.850  &   0.17  & 1   & 0.500  & 11  &  0.080 & 17  \\
1611$+$343  &  Qc  &  1.401  &  43.2  &  1.460  &   0.08  & 5   & 0.570  & 5   &  0.240 & 15  \\
1618$+$177  &  Ql  &  0.555  &  10.7  &  0.086  &   0.20  & 4   & 0.100  & 4   &  0.300 & 4   \\
1622$-$297  &  Q   &  0.815  &  43.2  &  2.355  &   0.06  & 5   & 0.400  & 5   &  0.080 & 17  \\
1633$+$382  &  Qc  &  1.814  &  22.2  &  1.553  &   0.07  & 5   & 0.200  & 5   &  0.250 & 15  \\
1637$+$826  &  G   &  0.023  &  10.7  &  0.670  &   0.20  & 4   & 0.300  & 4   &  0.300 & 4   \\
1641$+$399  &  Qc  &  0.593  &  22.0  &  6.900  &   0.30  & 4   & 0.490  & 11  &  0.660 & 4   \\
1642$+$690  &  Qc  &  0.751  &  15.0  &  1.180  &   0.05  & 1   & 0.380  & 11  &  0.145 & 18  \\
1652$+$398  &  BL  &  0.034  &  15.0  &  0.540  &   0.21  & 1   & 0.960  & 5   & 10.100 & 17  \\
1655$+$077  &  Qc  &  0.621  &  15.0  &  1.590  &   0.23  & 1   & 0.430  & 11  &  0.153 & 19  \\
1656$+$053  &  Qc  &  0.879  &  15.0  &  0.660  &   0.23  & 1   & 0.090  & 11  &  0.353 & 14  \\
1656$+$477  &  Qc  &  1.622  &  15.0  &  0.680  &   0.09  & 1   & 0.060  & 11  &  0.041 & 14  \\
1721$+$343  &  Ql  &  0.205  &  10.7  &  0.109  &   0.24  & 4   & 0.280  & 4   &  1.900 & 4   \\
1730$-$130  &  Qc  &  0.902  &  43.2  &  5.850  &   0.07  & 5   & 0.480  & 11  &  0.630 & 17  \\
1749$+$096  &  BL  &  0.320  &  15.0  &  5.550  &   0.05  & 1   & 0.150  & 11  &  0.150 & 15  \\
1749$+$701  &  BL  &  0.770  &  15.0  &  0.570  &   0.08  & 1   & 0.260  & 4   &  0.150 & 15  \\
1758$+$388  &  Qp  &  2.092  &  15.0  &  1.620  &   0.13  & 1   & 0.002  & 11  &  0.131 & 14  \\
1800$+$440  &  Qc  &  0.663  &  15.0  &  1.380  &   0.03  & 1   & 0.560  & 11  &  0.111 & 14  \\
1803$+$784  &  BL  &  0.684  &   5.0  &  1.436  &   0.20  & 4   & 0.004  & 4   &  0.160 & 4   \\
1807$+$698  &  BL  &  0.051  &  15.0  &  0.830  &   0.12  & 1   & 2.600  & 4   &  0.300 & 15  \\
1823$+$568  &  BL  &  0.664  &  15.0  &  2.140  &   0.12  & 1   & 0.120  & 4   &  0.200 & 4   \\
1828$+$487  &  Ql  &  0.692  &  15.0  &  1.300  &   0.07  & 1   & 0.380  & 11  &  0.344 & 19  \\
1830$+$285  &  Ql  &  0.594  &   8.5  &  0.380  &   0.24  & 8   & 0.130  & 12  &  0.310 & 6   \\
1845$+$797  &  Qc  &  0.057  &  15.0  &  0.300  &   0.13  & 1   & 0.600  & 11  &  5.470 & 15  \\
1921$-$293  &  Qc  &  0.352  &  15.0  &  0.410  &   0.19  & 1   & 0.190  & 11  &  1.060 & 15  \\
1928$+$738  &  Qc  &  0.302  &  15.0  &  2.580  &   0.11  & 1   & 0.600  & 4   &  0.550 & 4   \\
2007$+$776  &  BL  &  0.342  &   5.0  &  1.361  &   0.19  & 4   & 0.180  & 4   &  0.110 & 4   \\
2131$-$021  &  BL  &  1.285  &  15.0  &  1.150  &   0.13  & 1   & 0.120  & 11  &  0.050 & 15  \\
2134$+$004  &  Qp  &  1.932  &  15.0  &  2.020  &   0.15  & 1   & 0.020  & 11  &  0.260 & 15  \\
2136$+$141  &  Qc  &  2.427  &  15.0  &  2.040  &   0.12  & 1   & 0.020  & 11  &  0.120 & 23  \\
2144$+$092  &  Qc  &  1.113  &  15.0  &  0.550  &   0.07  & 1   & 0.030  & 11  &  0.035 & 13  \\
2145$+$067  &  Qc  &  0.999  &  15.0  &  7.970  &   0.11  & 1   & 0.030  & 11  &  0.360 & 16  \\
2200$+$420  &  BL  &  0.069  &  15.0  &  2.960  &   0.10  & 1   & 1.410  & 11  &  2.200 & 15  \\
2201$+$315  &  Qc  &  0.298  &  15.0  &  2.710  &   0.08  & 1   & 0.340  & 11  &  3.780 & 23  \\
2223$-$052  &  Qc  &  1.404  &  15.0  &  1.980  &   0.10  & 4   & 0.490  & 11  &  0.270 & 15  \\
\hline
\end{tabular}
\end{minipage}
\end{table*}

\newpage
\addtocounter{table}{-1}
\begin{table*}\centering\begin{minipage}{140mm}
\caption{$Continued.$}
\begin{tabular}{cccrccccrrr}
\hline\hline
 Source &
Type & z & $\nu_{\rm s}$ & $f_{\rm c}$ & $\theta_{\rm d}$ & Refs.
& $\mu_{\rm app}$ & Refs. & $f_{\rm 1keV}$ & Refs. \\
 &  &  & (GHz) &
(Jy) & (mas) &  & ($\rm mas~yr^{-1}$) &  & ($\rm \mu Jy$) &  \\
\hline
2230$+$114  &  Qc  &  1.037  &  43.2  &  2.428  &   0.04  & 5   & 0.500  & 4   &  0.730 & 15  \\
2234$+$282  &  Qc  &  0.795  &  15.0  &  0.460  &   0.32  & 1   & 0.120  & 11  &  0.050 & 16  \\
2243$-$123  &  Qc  &  0.630  &  15.0  &  1.920  &   0.17  & 1   & 0.290  & 11  &  0.279 & 24  \\
2251$+$158  &  Qc  &  0.859  &  43.2  &  2.015  &   0.05  & 5   & 0.530  & 5   &  1.370 & 15  \\
2345$-$167  &  Qc  &  0.576  &  15.0  &  1.460  &   0.10  & 1   & 0.030  & 11  &  0.180 & 16  \\
\hline
\end{tabular}
\begin{quote}
Column (1): IAU name; Column (2): classification of the source (Q=
quasars; Qc= core-dominated quasars; Ql= lobe-dominated quasars; Qp=
GHz peaked quasars; BL= BL Lac objects; G= radio galaxies); Column
(3): redshift z; Column (4): observation frequency $\nu_{\rm s}$ in
GHz; Column (5): core radio flux density $f_{\rm c}$ at frequency
$\nu_{\rm s}$; Column (6): VLBI core size $\theta_{\rm d}$ in mas;
Column (7): reference for the VLBI data; Column (8): the proper
motion $\mu_{\rm app}$; Column (9): reference for the proper motion;
Column (10): 1 keV X-ray flux density $f_{\rm 1keV}$ in $\mu$Jy;
Column (11): reference for the X-ray flux.

References: (1) Kovalev et al. (2005); (2) Brunthaler et al. (2000);
(3) Fey \& Charlot (1997); (4) J98; (5) Jorstad et al. (2001); (6)
Rokaki et al. (2003); (7) Lister (2001); (8) Fey \& Charlot (2000);
(9) Hough et al. (2002); (10) Fey, Clegg \& Fomalont (1996); (11)
Kellermann et al. (2004); (12) Vermeulen \& Cohen (1994); (13)
Siebert et al. (1998); (14) Brinkmann, Yuan \& Siebert (1997); (15)
Donato et al. (2001); (16) Ghisellini et al. (1993); (17) Comastri
et al. (1997); (18) Gambill et al. (2003); (19) Marshall et al.
(2005); (20) Reich et al. (2000); (21) Imanishi \& Ueno (1999); (22)
Galbiati et al. (2005); (23) Bloom et al. (1999); (24) Donato,
Sambruna \& Gliozzi (2005).
\end{quote}
\end{minipage}
\end{table*}

\begin{table*}\centering\begin{minipage}{140mm}
\caption{Derived Jet Parameters and Luminosity of the Sample.}
\begin{tabular}{crrrcccccc}
\hline\hline
 Source & $\theta$ &
$\Gamma$ & $\delta$ & $n_{1}$ & log $L_{\rm kin}$ & log $L_{\rm
BLR}$ &
Refs. & log $L_{\rm ext, 5 GHz}$ & Refs. \\
& (degree) &  &  & ($\rm cm^{-3}$) & ($\rm erg~s^{-1}$) & ($\rm
erg~s^{-1}$) &  & ($\rm
erg~s^{-1}$) &  \\
 \hline
0003$-$066  &     56.7  &     1.0  &     1.1 &    9.48E+04&  45.19   &  43.15 &  1     &  41.57  &   21   \\
0007$+$106  &     51.0  &     1.9  &     1.1 &    1.32E+04&  46.17   &  44.14 &  2     &  39.00  &   22    \\
0016$+$731  &      4.4  &    17.7  &    12.4 &    6.75E+04&  47.27   &  44.98 &  1     &  43.88  &   23  \\
0035$+$413  &     17.1  &    19.2  &     1.1 &    5.01E+06&  49.14   &  44.64 &  1     &  43.41  &   24   \\
0106$+$013  &      2.8  &    23.9  &    20.1 &    1.15E+05&  47.51   &  46.15 &  1     &  44.51  &   25    \\
0108$+$388  &     25.7  &     5.9  &     1.5 &    1.24E+05&  47.47   &    ... &  ...   &  41.65  &   26  \\
0112$-$017  &     23.4  &     1.7  &     2.3 &    9.52E+04&  46.97   &  45.26 &  1     &  43.38  &   25    \\
0133$+$207  &     18.0  &    28.6  &     0.7 &    2.09E+06&  48.77   &  45.02 &  1     &  43.26  &   25    \\
0133$+$476  &      4.5  &     3.7  &     6.7 &    4.42E+04&  46.96   &  44.47 &  1     &  41.85  &   21   \\
0153$+$744  &     14.5  &    12.3  &     2.3 &    2.23E+06&  48.78   &  46.14 &  1     &  42.79  &   23   \\
0202$+$149  &     12.1  &     6.6  &     4.5 &    1.63E+04&  46.60   &    ... &  ...   &  41.23  &   21  \\
0208$-$512  &      2.2  &    32.0  &    25.9 &    2.07E+04&  46.77   &  45.19 &  3     &  43.52$^{b}$  &   27   \\
0212$+$735  &      8.9  &     8.4  &     6.2 &    5.06E+05&  48.11   &  44.95 &  1     &  42.52  &   25    \\
0219$+$428  &      3.1  &    38.0  &    14.6 &    1.89E+04&  46.73   &    ... &  ...   &  43.06  &   28    \\
0234$+$285  &      4.8  &    20.6  &    10.5 &    1.05E+05&  47.47   &  45.26 &  1     &  42.61  &   25    \\
0235$+$164  &      2.4  &    66.5  &    14.9 &    1.37E+05&  47.60   &  43.86 &  1     &  42.36  &   25    \\
0316$+$413  &     88.9  &     1.3  &     0.8 &    4.76E+03&  45.23   &  42.70 &  4     &  40.64  &   29    \\
0333$+$321  &      3.3  &    27.5  &    15.4 &    1.01E+05&  47.46   &  45.94 &  5     &  42.90  &   5     \\
0336$-$019  &      2.5  &    19.7  &    22.5 &    9.73E+03&  46.43   &  45.00 &  1     &  42.94  &   25    \\
0415$+$379  &     21.9  &    17.3  &     0.8 &    2.09E+05&  47.76   &    ... &  ...   &  42.09  &   23   \\
0420$-$014  &      5.7  &    15.6  &     9.2 &    1.01E+05&  47.44   &  44.92 &  1     &  42.24  &   25    \\
0430$+$052  &      8.6  &     6.0  &     6.6 &    3.43E+02&  44.92   &  42.93 &  6     &  39.40  &   22    \\
0440$-$003  &      6.7  &    29.0  &     4.6 &    2.89E+05&  47.91   &  44.77 &  1     &  42.74  &   25    \\
0454$+$844  &     49.0  &     1.4  &     1.3 &    3.45E+03&  45.34   &    ... &  ...   &  40.44  &   30    \\
0458$-$020  &      3.1  &    13.8  &    17.7 &    4.42E+04&  47.08   &  45.32 &  1     &  44.39  &   25    \\
0528$+$134  &      2.6  &    38.7  &    19.0 &    3.22E+05&  47.97   &    ... &  ...   &  43.54  &   21  \\
0552$+$398  &     25.9  &     4.9  &     1.7 &    5.03E+06&  49.06   &  46.28 &  7     &  45.43  &   7,31  \\
0605$-$085  &     11.9  &    13.4  &     3.1 &    4.77E+05&  48.11   &  44.62 &  1     &  42.98  &   25    \\
0607$-$157  &      9.9  &     3.9  &     5.4 &    1.39E+04&  46.46   &  43.56 &  1     &  $<$41.95  &   25    \\
0615$+$820  &     40.0  &     2.6  &     1.3 &    1.02E+05&  47.22   &    ... &  ...   &  42.78  &   23  \\
0642$+$449  &      7.8  &     2.2  &     3.9 &    5.30E+05&  47.88   &  46.11 &  1     &  42.98  &   25    \\
0710$+$439  &     74.8  &     3.5  &     0.4 &    2.40E+06&  48.68   &    ... &  ...   &  42.87  &   26  \\
0716$+$714  &      4.8  &    49.5  &     5.5 &    6.91E+04&  47.30   &    ... &  ...   &  42.21  &   21   \\
0723$+$679  &     11.9  &    15.3  &     2.8 &    3.82E+05&  48.02   &  44.80 &  1     &  43.84  &   25    \\
0735$+$178  &      6.1  &    25.2  &     6.2 &    4.70E+04&  47.12   &    ... &  ...   &  40.27  &   30    \\
0736$+$017  &      3.3  &    12.3  &    16.5 &    1.05E+03&  45.45   &  44.18 &  1     &  40.74  &   25    \\
0738$+$313  &     20.8  &     2.7  &     2.8 &    2.64E+04&  46.66   &  45.78 &  1     &  42.25  &   25    \\
0745$+$241  &      9.6  &     8.6  &     5.6 &    1.20E+04&  46.49   &    ... &  ...   &  42.29  &   30    \\
0748$+$126  &      4.0  &    13.2  &    14.3 &    2.69E+04&  46.86   &  44.95 &  5     &  42.25  &   5     \\
0754$+$100  &      6.6  &    12.8  &     8.1 &    8.83E+03&  46.38   &    ... &  ...   &  40.39  &   30    \\
0804$+$499  &      6.9  &     8.8  &     8.3 &    6.67E+04&  47.24   &  45.39 &  1     &  42.04  &   25    \\
0808$+$019  &      1.9  &     9.6  &    17.3 &    1.11E+04&  46.46   &    ... &  ...   &  43.59  &   30    \\
0814$+$425  &      5.4  &     6.2  &     9.2 &    1.11E+03&  45.43   &  41.81 &  1     &  41.34  &   25    \\
0823$+$033  &      4.2  &    14.6  &    13.7 &    8.33E+03&  46.36   &  43.40 &  1     &  40.82  &   25    \\
0827$+$243  &      4.0  &    32.1  &    10.7 &    1.23E+05&  47.54   &  44.93 &  8     &  42.18  &   32    \\
0829$+$046  &      4.8  &    17.6  &    11.3 &    1.97E+03&  45.74   &    ... &  ...   &  40.97  &   30    \\
0836$+$710  &      4.8  &    29.3  &     8.5 &    1.50E+06&  48.63   &  46.43 &  1     &  43.78  &   25    \\
0850$+$581  &      8.7  &    37.6  &     2.2 &    2.02E+06&  48.76   &  45.66 &  1     &  43.78  &   25    \\
0851$+$202  &      6.3  &    14.3  &     8.3 &    1.87E+04&  46.71   &  43.60 &  1     &  40.48  &   25    \\
0859$-$140  &      3.9  &    16.8  &    14.7 &    4.13E+04&  47.06   &  45.74 &  1     &  43.45  &   25    \\
0906$+$015  &      6.3  &    12.2  &     8.7 &    7.05E+04&  47.28   &  45.11 &  1     &  42.53  &   25    \\
0906$+$430  &      1.8  &    11.0  &    19.5 &    2.95E+03&  45.90   &  43.35 &  1     &  43.86  &   25    \\
0917$+$449  &      3.7  &    13.1  &    15.2 &    8.70E+04&  47.37   &  45.21$^{a}$ &  9 &    44.66  &   33   \\
0917$+$624  &      2.5  &    16.5  &    21.8 &    1.91E+04&  46.72   &  45.06 &  1     &  42.14  &   25    \\
0923$+$392  &     15.9  &    43.4  &     0.6 &    1.77E+07&  49.71   &  45.78 &  1     &  43.45  &   25    \\
0945$+$408  &      2.4  &    22.7  &    23.8 &    1.55E+04&  46.64   &  45.59 &  1     &  43.15  &   25    \\
0953$+$254  &      8.4  &    22.4  &     3.8 &    1.39E+05&  47.59   &  44.97 &  1     &  41.85  &   25    \\
0954$+$658  &      6.8  &    10.3  &     8.3 &    4.17E+03&  46.04   &  42.63 &  1     &  41.32  &   25    \\
1012$+$232  &      4.9  &     9.3  &    11.3 &    5.70E+03&  46.17   &  45.16 &  10    &  42.73  &   34    \\
1015$+$359  &      7.6  &    15.2  &     6.1 &    1.07E+05&  47.47   &  45.98 &  9  &   $<$43.48 &   35   \\
1039$+$811  &      7.4  &     4.9  &     7.0 &    2.53E+04&  46.76   &    ... &  ...   &  42.67 &    23   \\
\hline
\end{tabular}
\end{minipage}
\end{table*}

\newpage
\addtocounter{table}{-1}
\begin{table*}\centering\begin{minipage}{140mm}
\caption{$Continued.$}
\begin{tabular}{crrrcccccc}
\hline\hline
 Source & $\theta$ &
$\Gamma$ & $\delta$ & $n_{1}$ & log $L_{\rm kin}$ & log $L_{\rm
BLR}$ &
Refs. & log $L_{\rm ext, 5 GHz}$ & Refs. \\
& (degree) &  &  & ($\rm cm^{-3}$) & ($\rm erg~s^{-1}$) & ($\rm
erg~s^{-1}$) &  & ($\rm
erg~s^{-1}$) &  \\
 \hline
1040$+$123  &     18.0  &    11.8  &     1.6 &    1.10E+06&  48.47   &  45.11 &  1     &  44.04  &   25    \\
1049$+$215  &      9.5  &    10.0  &     5.3 &    1.35E+05&  47.55   &    ... &  ...   &  44.00  &   36   \\
1055$+$018  &      1.9  &     5.5  &    10.7 &    2.80E+04&  46.82   &  44.53 &  1     &  43.13  &   25    \\
1055$+$201  &     10.0  &    15.9  &     3.7 &    2.19E+05&  47.78   &  46.09 &  11    &  44.31  &   30    \\
1101$+$384  &     29.3  &     3.3  &     1.8 &    1.67E+03&  45.51   &  41.40 &  12    &  39.78  &   28    \\
1127$-$145  &      3.5  &    69.7  &     7.2 &    9.81E+05&  48.45   &  45.77 &  1     &  43.21  &   25    \\
1128$+$385  &      6.6  &     2.1  &     3.7 &    7.96E+04&  47.01   &  46.26 &  9     &  42.26  &   30    \\
1137$+$660  &     45.7  &     5.1  &     0.6 &    6.98E+05&  48.21   &  45.83 &  1     &  43.94  &   25    \\
1150$+$812  &     15.1  &    11.2  &     2.3 &    6.81E+05&  48.26   &    ... &  ...   &  42.57  &   30    \\
1156$+$295  &      2.5  &    22.2  &    23.1 &    1.05E+04&  46.47   &  44.90 &  5     &  42.99  &   5     \\
1219$+$285  &     25.1  &     7.0  &     1.4 &    1.12E+04&  46.44   &  42.25 &  13    &  38.94  &   28    \\
1222$+$216  &      3.6  &    27.4  &    13.9 &    1.21E+04&  46.54   &  44.73 &  14    &  43.01  &   21   \\
1226$+$023  &      6.2  &    27.3  &     5.6 &    1.29E+05&  47.56   &  45.59 &  1     &  43.14  &   25    \\
1228$+$127  &     85.3  &     1.6  &     0.7 &    1.21E+02&  44.00   &    ... &  ...   &  40.84  &   23   \\
1253$-$055  &      4.1  &    10.4  &    13.5 &    3.83E+04&  47.01   &  44.64 &  1     &  43.46  &   25    \\
1302$-$102  &     14.3  &     6.3  &     3.7 &    1.33E+04&  46.51   &  44.91 &  1     &  41.83  &   25    \\
1308$+$326  &      2.2  &    45.3  &    22.5 &    4.41E+04&  47.10   &  45.12 &  1     &  42.56  &   25    \\
1334$-$127  &      0.5  &     9.8  &    19.4 &    1.06E+04&  46.45   &  44.18 &  1     &  42.67  &   25    \\
1345$+$125  &     75.3  &     6.4  &     0.2 &    5.13E+05&  48.10   &    ... &  ...   &  $<$41.71  & 35   \\
1406$-$076  &      2.2  &    61.1  &    18.1 &    1.01E+05&  47.46   &  45.47 &  15    &  $<$43.81  & 33   \\
1413$+$135  &      2.1  &    10.7  &    18.6 &    5.48E+02&  45.16   &    ... &  ...   &  $<$39.06  & 37   \\
1458$+$718  &      6.4  &    13.1  &     8.3 &    4.58E+04&  47.09   &  45.47 &  1     &  44.74  &   25    \\
1508$-$055  &      3.1  &    43.6  &    13.4 &    8.48E+04&  47.39   &  45.52 &  15   &  44.72 & 36   \\
1510$-$089  &      5.2  &    27.9  &     7.5 &    4.26E+04&  47.08   &  44.64 &  1     &  42.25  &   25    \\
1532$+$016  &      8.0  &    49.7  &     2.0 &    5.15E+06&  49.17   &  44.84 &  1     &  42.51  &   25    \\
1546$+$027  &      7.7  &     2.4  &     4.2 &    1.85E+04&  46.46   &  44.68 &  1     &  41.27  &   25    \\
1548$+$056  &     14.3  &     3.6  &     4.0 &    5.14E+04&  47.02   &    ... &  ...   &  44.27  &   36   \\
1606$+$106  &      3.2  &    42.8  &    12.5 &    1.57E+05&  47.65   &    ... &  ...   &  42.58  &   30    \\
1611$+$343  &      2.9  &    82.9  &     9.1 &    1.07E+06&  48.49   &  45.91 &  1     &  43.39  &   25    \\
1618$+$177  &     33.1  &    10.3  &     0.6 &    1.17E+06&  48.49   &  46.14 &  5     &  43.46  &   5     \\
1622$-$297  &      4.9  &    21.3  &     9.8 &    6.06E+04&  47.23   &    ... &  ...   &  $<$43.43  &  35   \\
1633$+$382  &      4.2  &    15.7  &    13.7 &    9.90E+04&  47.43   &  45.84 &  1     &  43.04  &   25    \\
1637$+$826  &     77.7  &     1.1  &     1.0 &    4.94E+02&  43.76   &    ... &  ...   &  40.04  &   23   \\
1641$+$399  &      6.1  &    29.5  &     5.4 &    5.91E+05&  48.23   &  45.47 &  1     &  43.05  &   25    \\
1642$+$690  &      2.5  &    16.7  &    21.7 &    6.14E+03&  46.23   &  43.86 &  1     &  42.98  &   25    \\
1652$+$398  &     45.0  &     4.2  &     0.8 &    1.24E+04&  46.43   &  41.36 &  12    &  39.43  &   28    \\
1655$+$077  &      6.8  &    27.7  &     4.7 &    2.28E+05&  47.81   &  43.62 &  5     &  42.76  &   5     \\
1656$+$053  &     24.2  &     7.9  &     1.3 &    1.12E+06&  48.45   &  46.26 &  5     &  42.78  &   5     \\
1656$+$477  &     11.1  &     4.5  &     5.1 &    5.54E+04&  47.09   &  45.76 &  16  &  $<$43.51 &   33   \\
1721$+$343  &     29.7  &    14.5  &     0.5 &    7.05E+05&  48.28   &  44.62 &  1     &  42.22  &   25    \\
1730$-$130  &      3.6  &    26.2  &    14.1 &    1.32E+05&  47.57   &  44.83 &  17    &  42.93  &   32    \\
1749$+$096  &      1.1  &     8.9  &    17.3 &    2.62E+03&  45.83   &    ... &  ...   &  39.74  &   28    \\
1749$+$701  &      6.8  &    11.8  &     8.0 &    2.29E+04&  46.79   &    ... &  ...   &  42.24  &   38   \\
1758$+$388  &     24.2  &     1.0  &     1.3 &    1.43E+06&  46.73   &    ... &  ...   &  42.44  &   39   \\
1800$+$440  &      0.9  &    28.2  &    46.7 &    1.91E+03&  45.73   &    ... &  ...   &  43.28  &   36 \\
1803$+$784  &     10.6  &     1.1  &     1.6 &    1.04E+05&  46.17   &  44.56 &  1     &  41.49  &   25    \\
1807$+$698  &      9.8  &    10.5  &     5.0 &    6.15E+02&  45.21   &  41.40 &  12    &  40.96  &   28    \\
1823$+$568  &     10.4  &     4.7  &     5.5 &    2.95E+04&  46.82   &  43.32 &  1     &  43.28  &   25    \\
1828$+$487  &      3.8  &    15.0  &    15.2 &    1.30E+04&  46.55   &  45.26 &  1     &  44.65  &   25    \\
1830$+$285  &     22.0  &     6.9  &     1.8 &    1.79E+05&  47.65   &  45.41 &  1     &  42.97  &   25    \\
1845$+$797  &     41.9  &     3.8  &     0.9 &    1.02E+04&  46.33   &  42.99 &  5     &  41.11  &   5     \\
1921$-$293  &     25.5  &     9.0  &     1.1 &    3.90E+05&  48.01   &  43.67 &  18   &  $<$43.35  &   32    \\
1928$+$738  &      5.3  &    11.5  &    10.8 &    6.88E+03&  46.26   &  45.18 &  1     &  42.02  &   25    \\
2007$+$776  &      5.2  &     5.1  &     8.4 &    2.66E+03&  45.79   &    ... &  ...   &  41.28  &   28    \\
2131$-$021  &      9.5  &     7.8  &     5.8 &    7.38E+04&  47.27   &  43.66 &  19    &  42.94  &   32    \\
2134$+$004  &     18.2  &     2.1  &     2.9 &    3.41E+05&  47.64   &  46.29 &  1     &  42.46  &   25    \\
2136$+$141  &     10.6  &     2.6  &     4.1 &    2.44E+05&  47.60   &  46.02 &  1     &  41.84  &   25    \\
2144$+$092  &     17.4  &     2.1  &     3.0 &    3.30E+04&  46.65   &    ... &  ...   &  42.97  &   30    \\
2145$+$067  &      3.9  &     3.6  &     6.6 &    8.31E+04&  47.23   &  45.79 &  1     &  42.14  &   25    \\
2200$+$420  &      5.8  &     7.0  &     9.3 &    6.27E+02&  45.19   &  42.38 &  20    &  39.84  &   28    \\
2201$+$315  &      5.2  &     7.1  &    10.0 &    8.23E+03&  46.31   &  45.46 &  1     &  43.05  &   25    \\
2223$-$052  &      2.1  &    33.1  &    26.4 &    4.53E+04&  47.11   &  45.62 &  1     &  43.59  &   25    \\
\hline
\end{tabular}
\end{minipage}
\end{table*}

\newpage
\addtocounter{table}{-1}
\begin{table*}\centering\begin{minipage}{140mm}
\caption{$Continued.$}
\begin{tabular}{crrrcccccc}
\hline\hline
 Source & $\theta$ &
$\Gamma$ & $\delta$ & $n_{1}$ & log $L_{\rm kin}$ & log $L_{\rm
BLR}$ &
Refs. & log $L_{\rm ext, 5 GHz}$ & Refs. \\
& (degree) &  &  & ($\rm cm^{-3}$) & ($\rm erg~s^{-1}$) & ($\rm
erg~s^{-1}$) &  & ($\rm
erg~s^{-1}$) &  \\
 \hline
2230$+$114  &      2.7  &    27.9  &    20.3 &    5.51E+04&  47.19   &  45.89 &  1     &  43.33  &   25    \\
2234$+$282  &     20.9  &    16.8  &     0.9 &    2.58E+06&  48.85   &  44.52 &  1     &  42.05  &   25    \\
2243$-$123  &      8.4  &    12.6  &     5.8 &    7.80E+04&  47.32   &  45.28 &  1     &  42.70  &   25    \\
2251$+$158  &      3.8  &    32.4  &    11.5 &    1.67E+05&  47.68   &  45.68 &  1     &  43.84  &   25    \\
2345$-$167  &     20.7  &     1.6  &     2.3 &    3.55E+04&  46.47   &  44.38 &  5     &  42.66  &   25    \\
\hline
\end{tabular}
\begin{quote}
Column (1): IAU name; Column (2): the viewing angle of jet $\theta$;
Column (3): the Lorentz factor $\Gamma$; Column (4): the Doppler
factor $\delta$; Column (5): the normalization factor of electron
energy distribution $n_{1}$; Column (6): the bulk kinetic power of
jet $L_{\rm kin}$; Column (7): the total luminosity in broad
emission lines $L_{\rm BLR}$, $^a$: also see Chen, Gu \& Cao (2009);
Column (8): the references for flux of broad emission lines used to
estimate $L_{\rm BLR}$; Column (9): the
radio extended 5 GHz luminosity $L_{\rm ext, 5 GHz}$, 
$^b$: from ATCA images; Column (10): the references for the radio
extended flux.

References: (1) Cao \& Jiang (1999); (2) Sergeev et al. (1999); (3)
Scarpa \& Falomo (1997); (4) Ho et al. (1997); (5) Liu, Jiang \& Gu
(2006); (6) Wang, Lu \& Zhou (1998); (7) Rokaki et al. (2003); (8)
Our unpublished measurements of Mg II line; (9) SDSS spectra; (10)
Brotherton (1996); (11) Kuraszkiewicz et al. (2004); (12) C97; (13)
Marcha et al. (1996); (14) Fan, Cao \& Gu (2006); (15) Wilkes
(1986); (16) Walsh \& Carswell (1982); (17) Cao (2000);  (18)
Jackson \& Browne (1991); (19) Rector \& Stocke (2001); (20) Corbett
et al. (1996); (21) Cooper, Lister \& Kochanczyk (2007); (22) Wills
\& Browne (1986); (23) Kharb \& Shastri (2004); (24) Gelfand et al.
(2005); (25) Cao \& Jiang (2001); (26) Vermeulen \& Cohen (1994);
(27) Marshall et al. (2005); (28) Perlman \& Stocke (1993); (29)
Pedlar et al. (1990); (30) Antonucci \& Ulvestad (1985); (31)
Stanghellini et al. (1990); (32) Browne \& Murphy (1987); (33)
Punsly (1995); (34) Saikia et al. (1990); (35) Ulvestad et al.
(1981); (36) Zhang \& Fan (2003); (37) Perlman et al. (1994); (38)
Wu et al. (2007); (39) Tinti et al. (2005).
\end{quote}\end{minipage}
\end{table*}

\end{document}